\documentclass[preprint2,natbib209]{aastex}    
\usepackage[latin1]{inputenc}
\usepackage{amsmath}
\usepackage{amsfonts}
\usepackage{amssymb}
\usepackage{graphics,graphicx}
\usepackage{units}

\begin{document}
\title{Sloshing Gas in the Core of the Most Luminous Galaxy Cluster RXJ1347.5-1145}
\author{
Ryan E. Johnson \altaffilmark{1,3} \email{johnsonr@denison.edu} \and
John ZuHone\altaffilmark{2,3} \email{jzuhone@cfa.harvard.edu} \and
Christine Jones\altaffilmark{3} \email{cjf@cfa.harvard.edu} \and
William R. Forman\altaffilmark{3} \email{wrf@cfa.harvard.edu} \and
Maxim Markevitch\altaffilmark{2,3} \email{maxim@head.cfa.harvard.edu}
}
\altaffiltext{1}{Department of Physics and Astronomy, Denison University, Granville, OH, 43023, USA}
\altaffiltext{2}{Astrophysics Science Division, Laboratory for High Energy Astrophysics, Code 662, NASA/Goddard Space Flight Center, Greenbelt, MD 20771, USA}
\altaffiltext{3}{Smithsonian Astrophysical Observatory, Harvard-Smithsonian Center for Astrophysics, Cambridge, MA 02138, USA}

\begin{abstract}
We present new constraints on the merger history of the most X-ray luminous cluster of galaxies, RXJ1347.5-1145, based on its unique multiwavelength morphology.  Our X-ray analysis confirms that the core gas is undergoing ``sloshing'' resulting from a prior, large-scale, gravitational perturbation.  In combination with multiwavelength observations, the sloshing gas points to the primary and secondary clusters having had at least two prior strong gravitational interactions.  The evidence supports a model in which the secondary subcluster with mass M=4.8$\pm2.4 \times$ 10$^{14}$ M$_{\odot}$ has previously ($\gtrsim$0.6 Gyr ago) passed by the primary cluster, and has now returned for a subsequent crossing where the subcluster's gas has been completely stripped from its dark matter halo.  RXJ1347 is a prime example of how core gas sloshing may be used to constrain the merger histories of galaxy clusters through multiwavelength analyses.  \\
\end{abstract}

\textit{Subject keywords}: galaxies: clusters: general; galaxies: interactions; Xrays: galaxies: clusters; galaxies: interactions

\section{Introduction}
Clusters of galaxies represent the end state in the hierarchical growth of the largest gravitationally bound structures in the universe.  Outside their cores, the (near) universal temperature and mass profiles of relaxed clusters make them ideal tools for constraining cosmological parameters (e.g. $w_{0}$ and $\sigma_{8}$) which govern the growth of structure in the universe \citep{V09}.  We see this growth of structure in large scale simulations \citep[e.g. Millennium I and II;][]{Mill1,Mill2}, however, there are relatively few corresponding observational constraints in place.  This is particularly true for matching the predicted rates at which massive clusters assemble with direct observational constraints on those rates.   For individual clusters, one should use empirically derived initial conditions in order to constrain their merger histories and attempt to reproduce the observations in hand, as was done recently for the Virgo Cluster \citep{Elke11}.  One such cluster for which there exists a wealth of observational data is the most luminous X-ray cluster, RXJ1347.5-1145 ($z$=0.451).  

High resolution hydrodynamic simulations predict the effects that multiple gravitational interactions will have on galaxy clusters.  In particular, the resulting morphology of the low-entropy cluster core gas \citep{AM06, Poole08, Zuhone10, Elke11}.  In this paper, we use the multiwavelength observations, along with simulations, to propose a merger history for RXJ1347 which explains the current morphology of the cluster in optical, X-ray, and millimeter Sunyaev-Zel'dovich (SZ) observations.  In the following section, we describe the phenomenon of core gas sloshing in galaxy cluster cores. In $\S$\ref{sec:rxj1347_s2}, we present a brief discussion of the extensive multiwavelength coverage and analysis of the cluster.  In $\S$\ref{sec:rxj1347_s3} we discuss our analysis of the \textit{Chandra} X-ray observations and the identification of the sloshing gas in the cluster core.  Finally, in $\S$\ref{sec:rxj1347_s4} we present new constraints on the merger history of the cluster based on the accumulated multiwavelength observations.

\subsection{Core Gas Sloshing}
\label{sec:slosh}
The model for core gas sloshing has been explored numerically in several works \citep{TH05,AM06,Zuhone10,Elke11}, and identified observationally in many galaxy groups  \citep{Randall09b,Machacek10} and clusters \citep{MM01, Churazov03, MV07, Owers09b, Johnson10}.  

The sloshing phenomenon begins when two group or cluster-sized halos cross nearby one another ($\lesssim$1 Mpc depending on their masses).  As seen from the reference frame of the more massive object (the primary cluster), the less massive object (the subcluster) falls in on either an elliptical or hyperbolic orbit (the exact orbit is irrelevant for the subcluster's first crossing).  As the subcluster falls in, the primary cluster (its dark matter halo and cluster gas) experiences a gravitational force towards the incoming object.  The subcluster reaches pericenter in its orbit, at which point the gravitational force is at its peak, and after which point the primary cluster reaches its maximum displacement from its initial position.  

The wake of the passing subcluster directs the ambient velocity field towards the primary cluster, while at the same time the primary cluster continues to move towards the subcluster.  This leads to ram pressure on the primary cluster gas, causing it to compress along the leading edge of its trajectory.  As the subcluster recedes, the velocity field around the primary cluster changes rapidly, removing the ram pressure, and resulting in a ram pressure ``slingshot'' \citep{Hallman04}, where the primary cluster gas moves quickly ahead of the potential minimum, causing a separation between the two.  It is this initial separation that leads to the subsequent production of large-scale asymmetries in the gas distribution.  

After its initial separation, the primary cluster gas density peak (its ``cool core'') falls back towards the potential minimum, which has itself been displaced from its initial position.  The density peak moves against the flow of its own trailing gas, resulting in ram pressure along its leading edge.  This ram pressure causes a significant ($>2x$) compression of the gas along the edge, revealing itself as a crescent-shaped edge in the X-ray surface brightness distribution of the core gas.  Upon reaching the apocenter of its oscillation about the potential minimum, the cool core reverses direction and again falls back towards the potential minimum against the flow of its own trailing gas, resulting in the formation of another ram pressure front.  These fronts have been dubbed ``cold fronts'' \citep{V01}, as opposed to shock fronts, because they delineate the edge of the cool core's oscillatory motion, with the coolest gas interior to the front and the warmer gas outside.  This oscillation and front production continues for long periods ($>$1 Gyr) in the core but is progressively damped with each crossing of the density peak through the potential minimum.  

The above describes the essential components of the radial motion of the density peak about the potential minimum.  However, because angular momentum is transferred from the subcluster in this interaction, the density peak undergoes azimuthal, as well as radial, oscillation about the potential minimum.  In all but a head-on merger ($b$=0), the displaced gas core does not pass directly \textit{through} the potential minimum, but rather undergoes a damped oscillation \textit{around} the potential minimum.  Whether due to a head-on or off-axis collision, the resulting oscillatory motion of the lowest entropy core gas within the gravitational potential is called ``sloshing'' \citep{MM01}.   

Despite the dependence of the long-term evolution of these cold fronts on merger parameters (cluster mass ratios, impact parameters, etc.) and the plasma properties (e.g., viscosities, magnetic fields, etc.), there are several qualities of these sloshing cold fronts that appear universal and are most relevant to RXJ1347.  First is that, based on the simulations, there appears to be a minimum time after subcluster pericenter for the first cold fronts to form.  This is because the dense gas core in the primary cluster must first slingshot past the cluster's potential minimum and will only then form the first cold front as it falls back towards the potential minimum.  Though the ability to detect the first sloshing cold front formed is limited by the simulation spatial resolution, over an order of magnitude in both mass ratio and impact parameters, simulations show the first cold front does not form until $\approx$0.3 Gyr after the closest approach of the subcluster \citep{AM06,Zuhone10,Elke11}.  Second, the location of the cold fronts reveals when, and from which direction, the sloshing disturbance originated.  The radial distance (projected or not) from the X-ray peak where the cold fronts are found always increases with the time since cold front formation.  That is to say that the sloshing cold fronts which formed first are always further from the oscillating core than fronts which form at a later time.  Also, simulations show that the first cold front always forms on the opposite side of the cluster as the location of the initial disturbance.  This is because the initial displacement of the primary cluster's gas core occurs on the same side as the pericenter of the disturbing object with the first cold front then forming as the core falls back from this initial displacement.  And finally, simulations show that the inclination of the clusters' merger axis with respect to our line of sight dictates whether the cold fronts will appear to us as a spiral inflow (for a merger in the plane of the sky), concentric crescent shapes (for a merger along the line of sight), or some combination of the two resulting from an inclined spiral inflow.  We discuss the implications of these for the history of RXJ1347 in $\S$\ref{sec:Interp}.

\section{Previous Observations of RXJ1347}
\label{sec:rxj1347_s2}
As the most X-ray luminous cluster known \citep[with $L_X$=6.0 $\times$ 10$^{45}$ erg s$^{-1}$ from 2 to 10 keV;][]{RASS}, RXJ1347 has been extensively studied in the radio, millimeter, submillimeter \citep{Gitti07b, Mason10}, in the optical via both weak and strong lensing \citep{Bradac08, Miranda08}, as well as through spectroscopy of member galaxies \citep{CK02, Lu10}, and of course in X-rays \citep{Allen02, Gitti07a}.  Indeed, one could argue that of those clusters beyond $z$=0.3, only the ``Bullet'' cluster (1E 0657-56; \citet{MM04}) has received more attention.  To establish the context and relevance of our interpretation, we review here some of the previous observations of RXJ1347.  We focus primarily on those contemporary works most relevant to our discussion. 

\subsection{Previous X-Ray Observations of RXJ1347}
\begin{figure*}[!th]
\center
\includegraphics[scale=0.6]{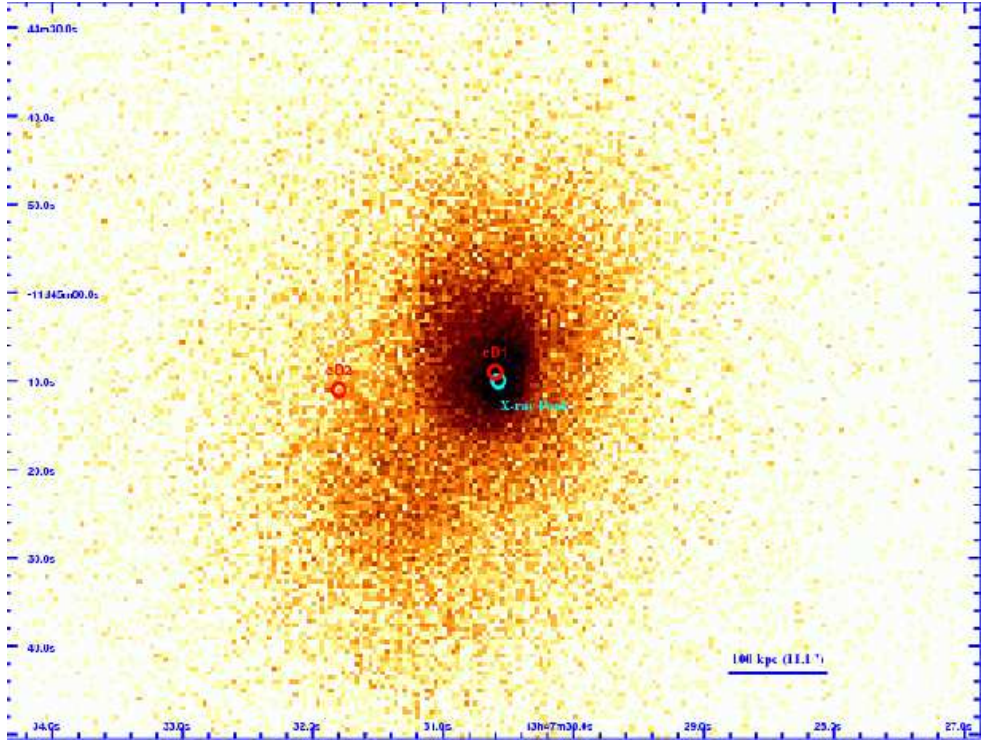}
\caption[Chandra image of RXJ1347]{\footnotesize{Combined 73 ks \textit{Chandra} image of RXJ1347.5-1145 from OBSIDs 506, 507, and 3592.  The image is shown in the 0.5-2.5 keV band and has been exposure corrected and background subtracted.  The surface brightness edges can be seen on either side of the emission peak.  The two small circles denote the locations of the primary and secondary cD galaxies (cD1 and cD2 respectively) and the small blue circle denotes the location of the X-ray peak.  In all images, north is up and east is to the left.  \textit{A color version of this figure is available in the online journal.} 
}}
\label{fig:rxj1347_f1}
\end{figure*}

RXJ1347 (see Fig.~\ref{fig:rxj1347_f1}) was first observed in X-rays with $ROSAT$ \citep{Schindler95}.  With the arcminute angular resolution of the $ROSAT$ PSPC, RXJ1347 appeared to be a relaxed system with a central cooling peak.  Since $ROSAT$, the cluster has been observed by nearly every X-ray satellite, including \textit{Suzaku} \citep{Ota08}, \textit{XMM-Newton} \citep{Gitti07a},  and \textit{Chandra} \citep{Allen02, Bradac08}.  
Conclusions regarding both the overall gas density, temperature, and metallicity are in agreement.  RXJ1347 exhibits a hot ($kT>$10 keV) intracluster medium (ICM), with an extension to the southeast.  Prior to the \textit{Chandra} observations \citep{Allen02}, there was little indication of this southeast structure (RXJ1347-SE hereafter) and, excluding this quadrant, the cluster gas distribution appears symmetric out to large scales ($>$1 Mpc), leading \citet{Allen02} to conclude that ``the cluster appears relatively relaxed.''  High-resolution imaging and spectral observations, with both \textit{XMM-Newton} and \textit{Chandra}, reveal a more complex picture with several features not typically seen in relaxed clusters in hydrostatic equilibrium, including a non-isotropic temperature distribution and two extremely hot regions ($kT>20$ keV) within 100 $h^{-1}$ kpc of the X-ray brightness peak (see Fig.~\ref{fig:rxj1347_f2}(c)).  

\subsection{Optical Observations of RXJ1347}
\label{sec:optical}
\citet{CK02} and \citet{Bradac08} show that the cluster hosts two cD galaxies, one 0.5$\arcsec$ (3.3 $h^{-1}$ kpc) NE of the X-ray peak (labeled cD1 in Fig.~\ref{fig:rxj1347_f1}) and the other, $\sim18\arcsec$ (118 $h^{-1}$ kpc) to the east, which has no detected X-ray counterpart (labeled cD2 in Fig.~\ref{fig:rxj1347_f1}).  The recessional velocities for these two galaxies differ by only $\sim200$ km s$^{-1}$ \citep{CK02,Lu10} which is much less than even the lowest estimate for the velocity dispersion of the cluster \citep[$\sigma_v$=910$\pm$130 km s$^{-1}$;][]{CK02}.  

A dedicated study was performed using 400 redshifts in the region surrounding RXJ1347 \citep{Lu10} which, along with the previous efforts by \citet{CK02}, brought the number of cluster members confirmed via redshifts to $\sim$140 galaxies.  \citet{Lu10} note several new structures in velocity space, located co-linearly in projection along a roughly NE/SW line through the cluster center \citep[see Fig. 1 in][]{Lu10}.  Among these galaxy overdensities is a significant (7$\sigma$)  peak in the redshift distribution corresponding to a cluster of galaxies approximately $\sim$7 $h^{-1}$ Mpc (projected) SW of the main cluster (referred to as RXJ1347-SW hereafter).   Although this structure appears to be much more distant than RXJ1347 (its mean redshift is $\sim$4000 km s$^{-1}$ larger), there is evidence that it is physically associated with RXJ1347, either through a previous interaction, which is unlikely given the separation, or by lying within the same cosmic filament as RXJ1347 \citep[see $\S$4.2 in][for a discussion]{Lu10}.  RXJ1347-SW lies outside both the \textit{Chandra} and \textit{XMM-Newton} observations for the cluster, so we use $ROSAT$ \citep{RASS} to place an upper limit on its luminosity of  
$L_X\lesssim7 \times 10^{44}$ erg  s$^{-1}$ which, using the $L_X/M_{gas,500}$ scaling relations in \citet{Zhang11}, predicts an upper limit to its mass (M$_{gas,500}\lesssim10^{14}$ M$_\odot$) that is in agreement with the mass estimate from \citet{Lu10}.  Another nearby overdensity of galaxies was noted in \citet{Bradac08} immediately to the SW of RXJ1347, coincident with a mass extension seen via weak+strong lensing.  We also do not detect this object in the \textit{Chandra} image and place an upper limit on its luminosity of $L_X<1.3 \times$ 10$^{42}$ erg s$^{-1}$ at the distance of RXJ1347.  

There is some disagreement among the different lensing analyses as to the presence of the second massive dark matter halo that would be associated with either cD2 or the gaseous subcluster.  In \citet{Bradac08} and \citet{Miranda08}, their projected mass density maps show an extension in the direction of the gaseous subcluster $\sim$30$\arcsec$ ($\sim$270 kpc) southeast of the X-ray brightness peak.  \citet{Miranda08} give an approximate mass for this structure comparable with the primary cluster mass ($\sim$10$^{15}$M$_{\odot}$).  On the other hand, the weak lensing analysis by \citet{Lu10} shows no such structure, and does not even show an extension of the weak lensing signal in that direction.  We discuss this further in terms of our proposed merger history for this cluster (see $\S$\ref{sec:rxj1347_s4}). 

\subsection{Radio Observations of RXJ1347}
\citet{Komatsu99} and \citet{Pointecouteau99,Pointecouteau01} made the
first measurements of the SZ effect in RXJ1347. The high angular resolution (20$\arcsec$ FWHM at 150 GHz) of the \citet{Komatsu01} observations showed a strong enhancement of the SZ signal 20$\arcsec$ southeast of the cluster center, at the same location as the surface brightness extension observed in X-rays (see Fig.~\ref{fig:rxj1347_f2} (a),(d)).  Since the SZ intensity is proportional to the integrated gas pressure along the line of sight, this strong SZ signal was interpreted as being due to hot gas, probably shock heated by the merger of the subcluster with the primary cluster. The presence of hot ($kT>15$ keV) gas in the southeast region has been reported from \textit{Chandra} \citep{Allen02, Miranda08}, \textit{XMM-Newton} \citep{Gitti07a} and $Suzaku$ observations \citep{Ota08}, and is confirmed here.

Recently, \citet{Mason10} and \citet{Korngut11} reported 90 GHz observations of RXJ1347 with 10$\arcsec$ angular resolution performed with the MUSTANG bolometer array on the Green Bank Telescope. In addition to confirming the enhanced SZ effect in the region of the subcluster merger, the MUSTANG observations (see Fig.~\ref{fig:rxj1347_f2}(d))  show a pronounced asymmetry in the projected pressure, with a high pressure ridge running north to south over $\sim0.5\arcmin$ ($\sim$197 $h^{-1}$ kpc) approximately 15$\arcsec$ ($\sim$99 $h^{-1}$ kpc) east of the X-ray peak.  The southern end of this ridge marks the peak of the SZ decrement, which \citet{Mason10} interpret as a shock between the primary cluster and infalling subcluster (RXJ1347-SE).  

We also note that \citet{Gitti07b} observed a radio mini-halo, consistent with the sloshing scenario \citep{Mazzotta08,Zuhone11b}.  It is $\sim$420 $h^{-1}$ kpc in extent and elongated to the southeast towards the location of RXJ1347-SE, with total flux density 55.3$\pm$0.6 mJy centered on the cluster core.  
The highest resolution Very Large Array image \citep[1.7$\arcsec \times 1.2\arcsec$ restoring beam;][]{Gitti07b} also shows extended emission in the core of RXJ1347, with a diffuse (i.e. point-source-subtracted) radio flux density of 25.2$\pm$0.3 mJy.  

\section{Our New X-ray Data Analysis of RXJ1347: ACIS-I Data Reduction, Imaging and Photometry}
\label{sec:rxj1347_s3}
\begin{figure*}[!th]
\center
\includegraphics[scale=0.8]{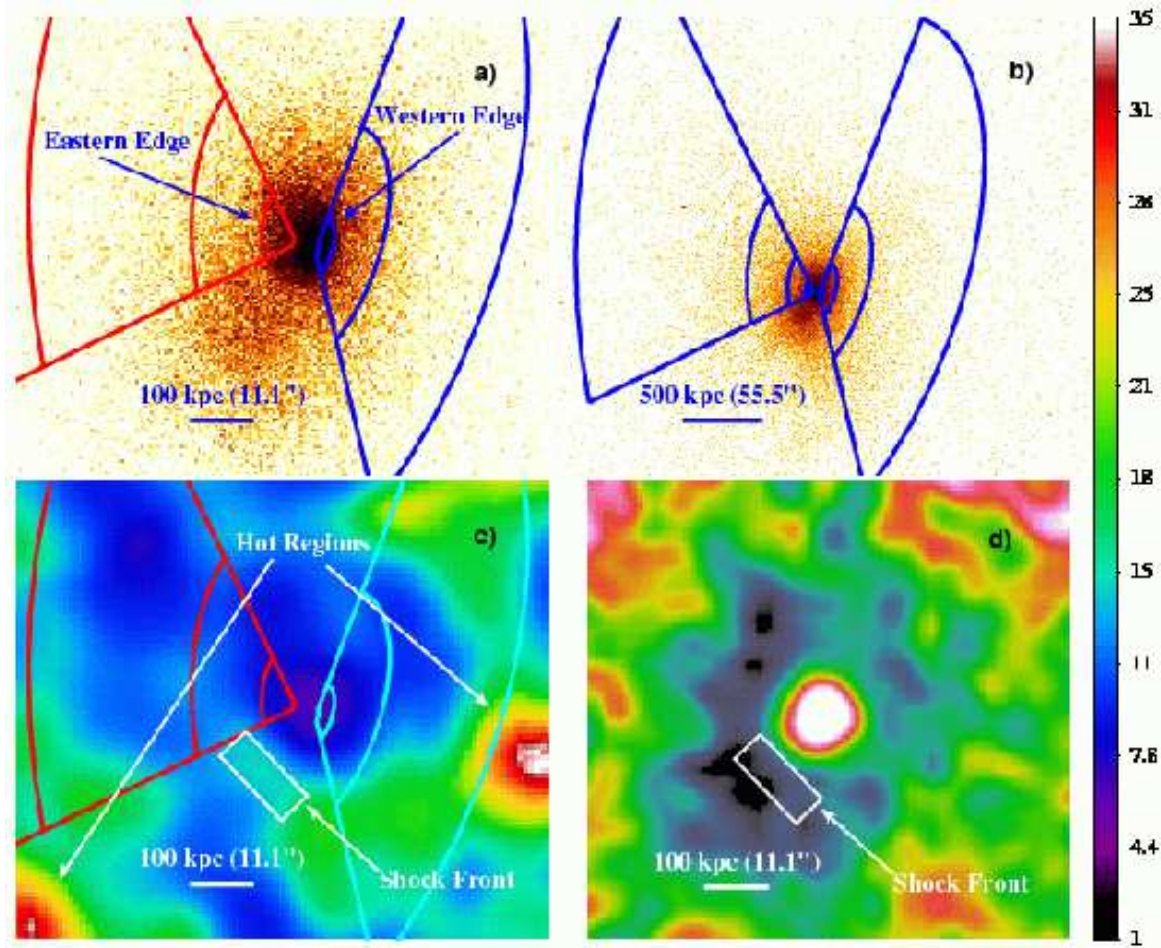}
\caption[Image of RXJ1347 showing annular regions]{\footnotesize{(a)  Same image as in Fig.~\ref{fig:rxj1347_f1} with annular sectors overlaid for surface brightness and spectral extraction.  (b) Same as in Fig.~\ref{fig:rxj1347_f1} with 2$\times$2 pixel binning, showing the spatial extent of the annular regions along with the large scale emission from the cluster.  (c) Temperature map produced as described in $\S$\ref{sec:tmap}.  (d) MUSTANG 90 GHz SZ map (Korngut et al. 2011) showing the SZ decrement ridge along the eastern side of RXJ1347.  Colors in (d) are of SZ decrement (lower value indicating stronger integrated pressure) measured lowest (black) to highest (white) with a logarithmic color stretch.  We also highlight the shock front which shows as a hot region in the temperature map and an SZ depression (indicating high integrated pressure) in the SZ map.  All images except (b) are shown at identical positions and scales. Color bar is linear scale in temperature going from cool (purple) to hot (white).  \textit{A color version of this figure is available in the online journal.}
}}
\label{fig:rxj1347_f2}
\end{figure*}

RXJ1347 was observed for a total of 80 ks over three \textit{Chandra} observations (ObsIDs 506, 507, and 3592).  The first two observations were performed with ACIS-S and the third with ACIS-I.  We reprocessed the level 1 event files as described in \citet{V05}.\footnote{All spectral extraction and spectral analysis were performed using CIAO ver. 4.3, the CALDB 4.4.1 and Sherpa \citep{Sherpa}.}  The background flare rejection was done using the ratio of the 2.5-7 keV and 9.0-12 keV bands \citep{HM06}, excluding regions with point sources and cluster emission.  This resulted in a combined $\sim$73 ks clean exposure, on which we performed the further image processing tasks described below.  Among the products created in this initial reduction are background images and exposure maps which incorporate the telescope aspect solution, vignetting, energy-dependent effective area, and position and energy-dependent detector efficiency.  The background-subtracted, exposure-corrected image used for all subsequent position and imaging analyses is shown in Fig.~\ref{fig:rxj1347_f1}, Fig.~\ref{fig:rxj1347_f2}(a) and (b) and Fig.~\ref{fig:rxj1347_f4}.  We detected point sources using the wavelet detection algorithm \textit{wvdecomp} outlined in \citet{V98}. The 30 point-source regions within 2.5$\arcmin$ of the X-ray peak were checked by eye and then excluded from all subsequent analyses.

Proper estimation of the X-ray background is critical for spectral analysis of the faint extended emission from clusters.  To measure this, we used the blank sky data sets from the \textit{Chandra} calibration database (CALDB ver. 4.4.1). The aspect solution for our observation was applied to the blank sky exposures, remapping the background events to our source coordinate frame. We then performed an overall background normalization by measuring the ratio of the count rates at high energies (9-12 keV, where the telescope's effective area is small) in our source images to that in our background images. For each of our spectral extraction regions, an identical region on the detector was chosen from the blank sky image to provide a background estimate.  Associated ancillary response files and redistribution matrix files were created using CIAO ver. 4.3.

\subsection{Gas Properties Across the Surface Brightness Edges}
\label{sec:SBprof}
From the source, background and exposure maps, we create surface brightness images with different pixel binning: full resolution (no binning) and block $n$ with $n \times n$ pixel binning (where $n=2,4,8$).  The full resolution and block two images reveal two sharp edges east and west of the X-ray brightness peak (see Figs.~\ref{fig:rxj1347_f1} and \ref{fig:rxj1347_f2}(a)). 

In RXJ1347, the presence of two surface brightness edges to the east and west indicates that the core gas may be sloshing, however it is not completely clear from which angle we are viewing it.  A spiral pattern is characteristic of sloshing core gas however, depending on the degree to which the merger is inclined with respect to the plane of the sky, this spiral pattern may appear elongated along one axis.  In RXJ1347, we may be viewing such an inclined spiral pattern elongated approximately along the N-S direction.  This can be seen in Fig. 1 beginning at a radius of $\sim10\arcsec$ away from the X-ray peak and position angle of 140$\degr$ (measured counter clockwise relative to west), and turning counterclockwise to a radius of $\sim5\arcsec$ with position angle 52$\degr$.  Such an elongation could indicate either a highly anisotropic ICM or gravitational potential into which the gas is expanding or, more likely, that the perturbing object came along a trajectory that was inclined with respect to the plane of the sky (see $\S$\ref{sec:Interp} for a discussion).   


Rather than the spiral pattern, if what we are viewing is instead the concentric crescent-shaped edges associated with a merger primarily along our line of sight, we then face the difficulty in explaining how both cD2 and the gaseous subcluster could be undergoing a merger with the primary cluster, since their projected distances would then suggest real distances which extend far outside the cluster core.  In this highly inclined merger scenario, we would also expect to see a significant difference in the redshifts of cD2 and cD1, which we do not..  We discuss the observational constraints and implications of both the line of sight and plane of the sky scenario in more detail in $\S$\ref{sec:Interp}.  

From the surface brightness images, radial sectors were chosen to encompass the east and west edges.  We measure the radial surface brightness profiles within these sectors and fit them with a two-component model for the gas density with a power law inside the surface brightness edge and a $\beta$-model outside with the form: 
\begin{equation}
\label{eqn:powbeta}
n_e(r) = n_1\left\{ \begin{array}{rcl}
& J_{n_e}(\frac{r}{r_j})^{\gamma_1} &\mbox{ for  $r \leq r_j$}\\
& [1+(\frac{r}{r_c})^{^2}]^{-\frac{3\beta}{2}} &\mbox{ for  $r>r_j$}\\
       \end{array} \right.
\end{equation}
where $r_j$ is the radius of the surface brightness edge, $\gamma_1$ is the power law slope interior to the edge,  $r_c$ is the core radius of the $\beta$-model, $n_1$ combines the model normalization with the emission measure normalization for each model in their respective regions, and $J_{n_e}$ is the magnitude of the discontinuous density jump at the edge. 

The square of this density model is integrated along the line of sight and compared to the observed surface brightness profile using a chi-squared minimization.  Due to the temperature dependence on the emissivity of the gas, and thus the surface brightness, this is an iterative process, where we first fit the surface brightness assuming a uniform temperature.  Then, upon deprojecting the temperature profile using the resulting density model (see $\S$\ref{sec:specfit}), we use the deprojected temperature profile to again fit for the emissivity-corrected surface brightness.  Convergence is obtained typically after only one such iteration.

\begin{figure*}[!th]
\center
\includegraphics[scale=0.60,angle=270]{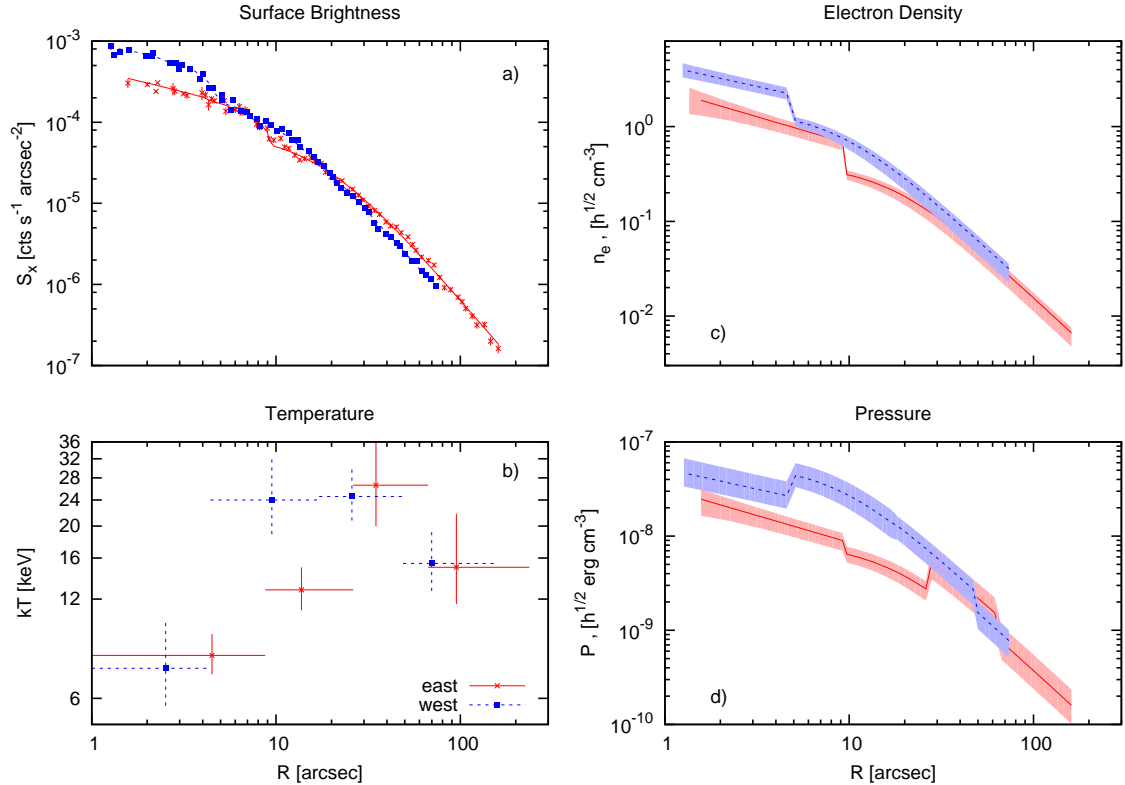}
\caption[Radial profiles for the cold fronts in RXJ1347]{\footnotesize{Radial quantities plotted for east and west regions shown in Fig.~\ref{fig:rxj1347_f2}(a) and (b).  (a) Radial surface brightness, (b) deprojected temperature, (c) electron density, and (d) pressure profiles for the regions east (solid x's, light (red) shading) and west (open dashed squares, dark (blue) shading). Spectral extraction and fitting are described in $\S$\ref{sec:specfit}.  Shadings in (c) and (d) are from the 90$\%$ confidence intervals on the model parameters.  \textit{A color version of this figure is available in the online journal.}
\label{fig:rxj1347_radial}
}}
\end{figure*}

We plot the resulting radial profile for the eastern surface brightness edge (red points, labeled ``east'' in Fig.~\ref{fig:rxj1347_radial}(a)), and find that the surface brightness profile across this feature may be adequately represented by Eq.~\ref{eqn:powbeta} with a jump of J$_{n_e}$=2.2$\pm0.6$ (see Table~\ref{table:rxj1347_edge_t}) at the radius of discontinuity ($r\sim9.2\arcsec\sim60 \ h^{-1}$ kpc).  The density model is plotted in Fig.~\ref{fig:rxj1347_radial}c.  
As discussed in $\S$\ref{sec:slosh}, given sufficient time since the initial sloshing disturbance, a surface brightness discontinuity on one side of the cluster should be accompanied by a second discontinuity on the opposite side at a different radius \citep[][; and $\S$\ref{sec:slosh} in this work]{AM06}.  Indeed, the western surface brightness jump lies $\sim4.6\arcsec$ ($\sim30 \ h^{-1}$ kpc) from the X-ray peak, closer than the eastern edge which lies $\sim9.2\arcsec$ ($\sim60 \ h^{-1}$ kpc) from the X-ray peak. Surface brightness and density plots are shown in Fig.~\ref{fig:rxj1347_radial}(a) and (c) (blue points, labeled ``west'').  That it is closer to the X-ray peak indicates that the western edge is the younger of the two, resulting from the most recent passage of the core gas across the cluster potential minimum.  This feature is also well fit by a two component density model, with a power law inside the edge, a $\beta$-model outside (Eq.~(\ref{eqn:powbeta})) and jump magnitude $J_{n_e}=2.0^{+0.6}_{-0.5}$.

\subsection{Temperature Maps}
\label{sec:tmap}
In Figure~\ref{fig:rxj1347_f2}c we present a temperature map for RXJ1347, also shown in \citet{Zuhone10}.
Temperature maps reveal interesting structures and inhomogeneities in the cluster gas which might otherwise be overlooked by only examining the X-ray surface brightness distribution. 

To generate the map, we use the method of \citet{MM00}, in which a series of 6 images in non-overlapping energy bands (0.5-1.5, 1.5-2.3, 2.3-3.5, 3.5-4.5, 4.5-6.5, 6.5-9.5 keV) are created along with corresponding exposure maps.  Each surface brightness image is then smoothed identically with a variable length scale Gaussian kernel, such that the smoothing scale at each pixel decreases roughly as the square root of the number of counts in that pixel (out to a brightness of $\sim$0.02 counts arcsec$^2$, below which the smoothing scale is constant).  This method has the effect of applying a smaller smoothing scale in areas of bright cluster emission and a larger smoothing scale in the low surface brightness regions, resulting in approximately the same relative statistical accuracy across the interesting bright regions of the cluster.  

The surface brightness images are then used to fit a \textsc{mekal} model spectrum for each image pixel, binned to those same energy bands, and a chi-squared minimization is performed allowing only the model temperature and normalization to vary.  Because this method does not assume any a priori shape to the spectral region, as opposed to tessellation or contour binning methods \citep[e.g.,][]{Sanders06}, we may be confident that any features shown, particularly near the center, reflect an accurate (projected) morphology of the cluster gas. 

The temperature map shows large-scale inhomogeneities of the cluster gas, indicative of an unrelaxed cluster.  Of particular note in this map are the cluster cool core, the presence of the eastern and western cold fronts seen in the surface brightness image, two very hot ($kT>$25 keV) regions, and the shock front between the primary and sub-clusters, also seen in the SZ map (Figure~\ref{fig:rxj1347_f2}(d)).  

\subsection{Spectral Extraction and Fitting}
\label{sec:specfit}
After fitting the X-ray surface brightness across the edges, we extract spectra in these same sectors (see Fig.~\ref{fig:rxj1347_f2}(a) and (b)).  The spectral bins are necessarily wider than the surface brightness bins as we require at least 2000 source counts (background subtracted) in the 0.8-7.0 keV range per spectral bin to constrain a thermal model with small enough uncertainties to identify the temperature jumps predicted at the cold front edges.  Spectral extraction and fitting were performed in an analogous way to our previous work \citep{Johnson10}, using the CIAO suite of tools. Spectra in each spatial region were binned with a minimum of 30 counts per energy channel and the models were fit over the range from 0.8 to 7.0 keV.  For each spectrum, we fit a thermal \textsc{mekal} model convolved with an intergalactic absorption model (xsmekal * xszwabs). The cluster's redshift was fixed at the median redshift of its cluster members ($z$=0.451) determined by \citet{Lu10} with an absorbing hydrogen column density $N_H=4.85 \times 10^{20}$ cm$^{-2}$ \citep{DL90} at the cluster position.  

We account for the gas lying in projection along the line of sight by using a variation of the algorithms \textit{projct} for \textit{XSpec} and \textit{deproject} for \textit{Sherpa}. Our algorithm takes a series of radially extracted spectra, fits a model to the outermost bin first and then moves inward. For each inward step, an additional model is added to the fit, whose gas temperature is fixed at the values obtained from the outer bins. The difference between \textit{deproject}, \textit{projct}, and our method lies in our separate derivation of the three-dimensional gas density rather than leaving the density as a free parameter in the model fit. The density for each radial bin is calculated as described in $\S$\ref{sec:SBprof} and the ratio of the emission measure in the projected region to that in the emission region is used as a normalization constraint for each of the projected regions. We present these radial quantities in Table~\ref{table:rxj1347_t1}. 

It should be noted that due to the small effective area of \textit{Chandra} at high energies, it is difficult to constrain high temperatures ($\gtrsim20$ keV) with confidence, even with high numbers of counts.  This is evidenced in the 90$\%$ confidence limits on the temperatures in the hottest radial bins (see Table~\ref{table:rxj1347_t1}).  Despite this limitation, we do find that the eastern and western edges identified in the surface brightness images are indeed cold fronts, as expected from the gas sloshing model \citep{AM06}.  From the deprojected temperature and density profiles, we are able to calculate the pressure profile across each edge which, in the absence of high radial velocities, should be approximately continuous \citep{AM06}.  In order to produce this profile we multiplied the mean temperature by the mean density in each spectral bin.  The mean pressure profile does not appear very smooth, however this is due primarily to the large spectral bins required to constrain the temperatures, and the large uncertainties in those temperatures.  Deeper X-ray observations are necessary to decrease the uncertainties in the spatial temperature distribution.  We note, however, that even with the coarse temperature resolution, including both the temperature and density uncertainties, each of the profiles is continuous at the 90$\%$ confidence level (see Fig.~\ref{fig:rxj1347_radial}(d)).

\begin{deluxetable}{cccc}
\center
\tablecaption{Radial Temperatures for RXJ1347}
\tablewidth{0pt}
\tabletypesize{\footnotesize}
\tablehead{
\colhead{Region} & \colhead{R} & \colhead{kT} & \colhead{$\chi^2$ (dof)}\\
& \colhead{[$\arcsec$]} & \colhead{[keV]} & 
} 
\startdata
West & 70.0 & 15.4$^{+4.1}_{-2.7}$ & 191.7 (197)\\
 & 25.8 & 24.6$^{+5.7}_{-3.8}$ & 230.4 (297)\\
 & 9.5 & 24.0$^{+8.2}_{-5.1}$ & 195.2 (257) \\
  & 2.5 & 6.4$^{+2.9}_{-1.7}$ & 87.0 (108)\\
\tableline 
East & 95.2 & 15.0$^{+6.8}_{-3.4}$ & 94.0 (80)\\
 & 34.9 & 26.8$^{+11.6}_{-6.8}$ &  110.7 (136)\\
 & 13.7 & 12.5$^{+2.2}_{-1.7}$ & 191.2 (214)\\
 & 4.5 & 8.1$^{+1.3}_{-1.0}$ & 112.9 (160)\\
\enddata
\tablecomments{\footnotesize{MEKAL model temperatures for the regions shown in Fig.~\ref{fig:rxj1347_f2}(a) and (b).  \textit{(Col.~1)} Spectral extraction region identifier. \textit{(Col.~2)} emission-weighted radius of spectral bin. \textit{(Col.~3)} deprojected MEKAL model temperature. Errors on $kT$ are the 90$\%$ confidence interval, $\Delta\chi^2$=2.71, for one interesting parameter). \textit{(Col.~4)} $\chi^2$ for model and number of degrees of freedom.}}
\label{table:rxj1347_t1}
\end{deluxetable}

\begin{deluxetable}{ccccc}
\center
\tablecaption{Edge Properties for RXJ1347}
\tablewidth{0pt}
\tabletypesize{\footnotesize}
\tablehead{
\colhead{Cluster ID} & \colhead{Edge Radius [$\arcsec$]} & \colhead{$\frac{T_2}{T_1}$} & \colhead{$\frac{n_{e1}}{n_{e2}}$} & \colhead{$\frac{P_1}{P_2}$}
} 
\startdata
East & 9.2$^{+0.5}_{-0.5}$ & 1.58$^{+0.53}_{-0.40}$ & 2.22$^{+0.60}_{-0.58}$ &  1.40$^{+0.65}_{-0.50}$\\
West & 4.6$^{+0.2}_{-0.2}$ & 3.24$^{+2.41}_{-1.41}$ & 1.98$^{+0.61}_{-0.45}$ & 0.61$^{+0.54}_{-0.28}$\\
\enddata
\tablecomments{\footnotesize{Values in each column are the ratios of various gas quantities (temperature, gas density and gas pressure) across the eastern and western surface brightness edges in RXJ1347 (see Fig.~\ref{fig:rxj1347_f1}), where $N_1$ represents the value of the gas property $N$ inside the front ($r<r_{edge}$) and $N_2$ is the value outside the front ($r>r_{edge}$).   The errors presented on the ratios are based on the 90$\%$ confidence limits from those quantities. 
}}
\label{table:rxj1347_edge_t}
\end{deluxetable}

\subsection{The Gaseous Subcluster}
\label{sec:rxj1347_subcluster}
\begin{figure*}[!th]
\center
\includegraphics[scale=0.6]{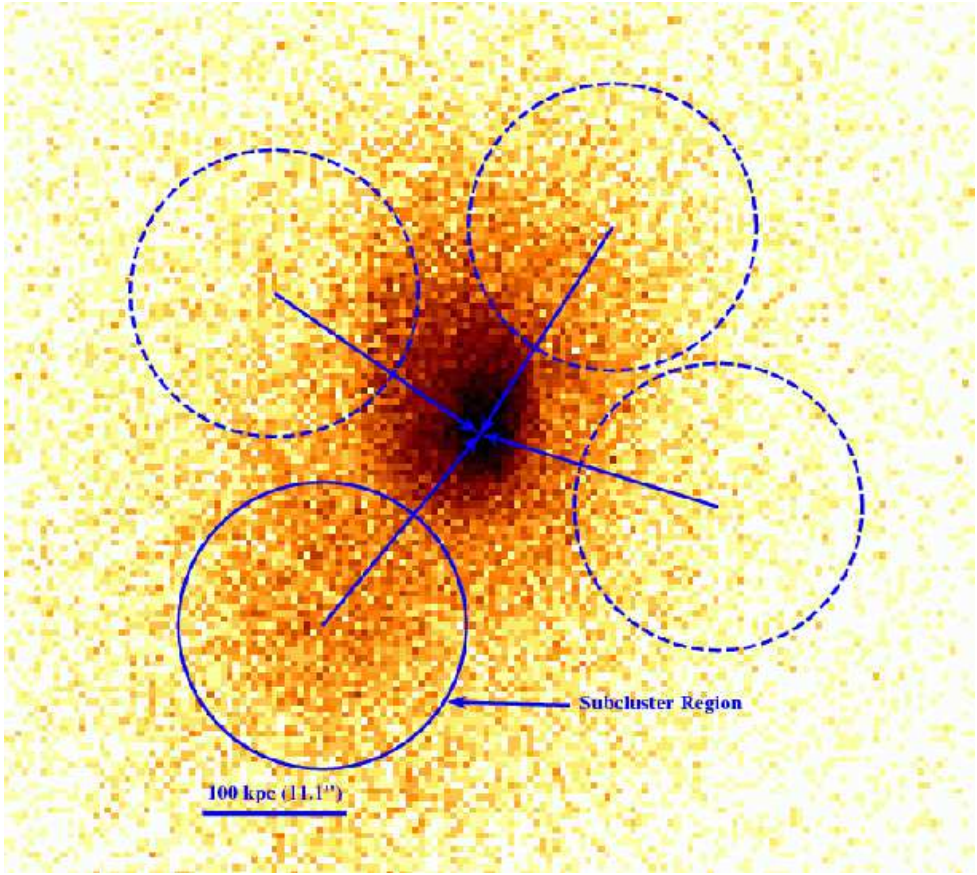}
\caption[Image of RXJ1347 showing regions use to estimate subcluster emission]{\footnotesize{Surface brightness image as in Fig.~\ref{fig:rxj1347_f1}, showing the regions used for estimating the subcluster emission.  Each circular region is identical in radius, with their centers equidistant from the cluster X-ray peak (shown with vectors pointing from each circle).  The source region used for the subcluster emission is labeled (blue, solid), whereas the other circles (blue-dashed) are used to represent the cluster emission at that same radius.  \textit{A color version of this figure is available in the online journal.}
}}
\label{fig:rxj1347_f4}
\end{figure*}

The gaseous subcluster to the southeast of the X-ray core lies within the halo of RXJ1347 (see Fig.~\ref{fig:rxj1347_f1}), making a study of its spectral properties problematic.  We isolate its emission by selecting a circular region in the surface brightness image containing its emission (see Fig.~\ref{fig:rxj1347_f4}) along with three other identically sized regions at the same distance from, but at different position angles with respect to, the X-ray peak.  The flux we measure in the region surrounding the subcluster includes some cluster emission, which we measure as $f_{x,sub+clust}=3.8\pm0.5 \times 10^{-13}$ erg s$^{-1}$ cm$^{-2}$ and compare with the mean flux in the other three regions ($\bar{f}_{x,other}=2.3\pm0.4 \times 10^{-13}$ erg s$^{-1}$ cm$^{-2}$).\footnote{All fluxes here computed in the 0.5-2.5 keV energy band.  Errors are systematic variation of 1$\sigma$ based on the three regions surrounding the cluster.}   So our estimate of the subcluster flux is the difference of these, or ${f}_{x,sub}=1.5\pm0.9 \times 10^{-13}$ erg s$^{-1}$ cm$^{-2}$.  This subcluster flux translates to a subcluster luminosity (at the distance of RXJ1347) of $L_{x,sub}=1.1\pm0.6 \times 10^{44}$ erg s$^{-1}$.  Using the L-M$_{gas}$ scaling relations from \citet{Zhang11},\footnote{There are separate scaling relations for both disturbed and undisturbed systems, and we use the former.} this corresponds to a subcluster gas mass of M$_{gas,sub} \sim 4.8\pm2.5 \times 10^{13} M_{\odot}$ within r$_{500}$.  Assuming a core gas mass fraction $f_{gas}$=0.1 \citep{V06}, this gives a mass for the subcluster of M$_{500,sub}=4.8\pm2.5 \times 10^{14} M_{\odot}$.  Given the merger scenario we propose in $\S$\ref{sec:rxj1347_s4}, the subcluster would have likely lost some portion of its mass during its first pericentric core passage.  This suggests that our mass estimate is probably a lower limit.   

\section{Discussion: The Merger History of RXJ1347}
\label{sec:rxj1347_s4}
While previous works agree that RXJ1347 appears to have undergone a significant merger \citep{Gitti07b,Lu10}, there is little consensus on the details.  The cluster has a complex morphology at nearly every wavelength, and there is no single merger scenario thus far presented which can explain the unique characteristics of this cluster.  Specifically, the key observables that must be explained by an acceptable merger model are as follows: 
\begin{itemize}
\item The presence of two sloshing edges: one at $\sim9.2\arcsec$ ($\sim$60 $h^{-1}$ kpc) east of the X-ray peak, and a second $\sim4.6\arcsec$ ($\sim$30 $h^{-1}$ kpc) to the west of the X-ray peak.  
\item A very hot region located $\sim17\arcsec$ ($\sim$113 $h^{-1}$ kpc) to the SE of the primary cluster. 
\item A second cD galaxy with no detected X-ray gas atmosphere located $\sim18\arcsec$ (118 $h^{-1}$ kpc) directly east of the primary cluster and $\sim11\arcsec$ (72 $h^{-1}$ kpc) north of the gaseous subcluster.  
\end{itemize}

\subsection{Previous Merger Scenario}
\label{sec:prev_scen}
The presence of the high pressure ridge along the east side of the cluster (see Fig.~\ref{fig:rxj1347_f2}(d)), connecting the second bright central galaxy (cD2 in Fig.~\ref{fig:rxj1347_f1}) to the gaseous subcluster, suggests that these objects are all related.  As noted in \citet{Mason10}, these features are both consistent with cD2 having moved north (in projection) along the east side of RXJ1347, with the gaseous subcluster having been either stripped or shock heated (or both) as a result of this passage.  This is the simplest model that agreed, until now, with all of the observations.  One problem with this model is that sloshing edges are produced well after the passage of the subcluster, which does not agree with the current location of either cD2 or the gaseous subcluster. 

\subsection{Sloshing Constraints on the Merger History  - A Second Crossing}
\label{sec:second_pass}
As discussed in $\S$\ref{sec:slosh}, simulations of gas sloshing predict a minimum time ($\approx$0.3 Gyr) should elapse between the first pericentric crossing  ($t_{per}$) of the perturbing cluster and the formation of the first sloshing cold front, irrespective of the merger axis inclination with respect to the line of sight.  For the subcluster\footnote{Here we make special distinction between the gaseous subcluster, RXJ1347-SE, and the subcluster in general, comprising cD2, the gaseous subcluster, and their combined dark matter halo.} to have perturbed RXJ1347, it should now be a minimum distance away from pericenter for us to see sloshing cold fronts in the core.  For example, if we assume a reasonable value for the subcluster's average velocity since $t_{per}$, say 1000 km s$^{-1}$, then it should be at least $\sim$300 kpc away in 0.3 Gyr.  This minimum distance is further constrained by the observation of a second sloshing cold front, which occurs only after the density peak has fallen past the potential minimum, reached apocenter, and has then fallen back again.\footnote{This is assuming a near plane of the sky viewing angle.}  Simulations show that this second front forms, at the earliest, $\sim$0.6 Gyr after $t_{per}$, increasing the predicted separation distance between pericenter and the current location of the subcluster to $\sim$600 kpc.  This is considerably larger than the observed (projected) separation distance ($\sim$160 kpc) between RXJ1347 and the location of cD2 (which should also be the location of the subcluster, as both the dark matter and the cD galaxy pass essentially collisionlessly through the envelope of RXJ1347).  This discrepancy indicates that the simple scenario where the subcluster is merging for the first time is not sufficient.  We therefore propose a new scenario, which we will refer to as the ``second crossing'' scenario.

\subsection{A New Merger History For RXJ1347}
\label{sec:Interp}
The simple interpretation of RXJ1347 now experiencing the first crossing of the subcluster does not correctly predict the presence of the sloshing cold fronts observed in the cluster core.  It is also inconsistent with the simulation result that the primary cluster's core gas does not separate from its potential minimum until after the first passage of the subcluster.  We propose instead that the subcluster first passed on the west side of RXJ1347, moving in a (projected) southerly direction, initiated core gas sloshing upon reaching pericenter in its orbit, and has now returned for a second pass, this time from the south, and much closer to the core.  During this second passage, the subcluster had its core gas stripped as it approached pericenter, and is now near this point in its orbit.  This  ``second crossing'' scenario addresses each of the three cluster observables noted in $\S$\ref{sec:rxj1347_s4}.

\textit{The presence and location of two sloshing cold fronts around the primary cluster's core.}  By allowing the core gas sloshing to be initiated on the first subcluster passage, we solve the discrepancy between the minimum time since pericenter ($\sim$0.6 Gyr) and the current location of the subcluster.  In this scenario, we require that the time between pericentric crossings of the subcluster be larger than 0.6 Gyr, as simulations predict is the case \citep[cf.][]{TH05, AM06, Zuhone10, Elke11}.  Simulations also predict that the first cold front should be produced on the opposite side of the cluster from the perturber's pericentric location.  For RXJ1347, we propose that the initial subcluster pericentric crossing was on the west side of the cluster.  Thus the first cold front should be on the east side.  Our measurements of the cold front radii (see Table~\ref{table:rxj1347_edge_t}) indicate that the eastern cold front is farther from the X-ray peak (9.2$\arcsec$) than is the western cold front (4.6$\arcsec$), i.e. it was formed earlier, in agreement with the ``second crossing'' scenario.  We also note that the lowest mass subcluster that has been simulated to produce sloshing is of order 10$^{13}$  $M_{\odot}$ \citep{Elke11}.  As shown in $\S$\ref{sec:rxj1347_subcluster}, our estimate for the subcluster gas mass alone is $M_{gas}\sim 10^{13}$ $M_{\odot}$, putting its total mass $M_T\sim10^{14}$ $M_{\odot}$.  This implies the primary cluster is of order 10 times the mass of the subcluster, which is well above the mass ratio limit required to initiate sloshing \citep{Zuhone10}. 

\textit{A very hot region located between the primary and sub- cluster, and the current location of cD2 and the gaseous subcluster.}  Shocked gas between the primary cluster and the gaseous subcluster is predicted in all merger simulations for which the subcluster possesses gas of its own.  This is because, being initially bound to the subcluster, the subcluster gas velocity exceeds the sound speed of the main cluster as it approaches pericenter in its orbit.  The subcluster undergoing a second crossing will not change this.  The shock from the subcluster's first crossing would have passed through the primary cluster's core already, likely contributing to the initial displacement of the RXJ1347's gas core from the potential minimum \citep[][]{Churazov03}.  Before the subcluster reached apocenter on its first pass, this shock would have traveled well beyond the cluster core, so that there would be little evidence of the front during the subcluster's second crossing.  

As the subcluster returned for its second crossing, this time from the south, its trajectory brought it closer to the core of RXJ1347, where it is now near pericenter, at $\sim18\arcsec$ (118 $h^{-1}$ kpc) east of the primary cluster center.  At this closer pericentric distance, the combination of the subcluster's high velocity and the ICM density of RXJ1347 increased the ram pressure on the subcluster's core gas enough to strip the gas from the dark matter halo. The gaseous subcluster is still traveling supersonically so we see a shock front between it and the core of RXJ1347.  \citet[][see their Fig.~4]{Bradac08} produce the highest resolution lensing map of the region around RXJ1347 and show that the projected mass density is elongated towards the region in between cD2 and the gaseous subcluster.  That the mass density elongation does not lie directly on the gaseous subcluster, or on cD2, indicates that some fraction of the projected mass density is lying north of the subcluster in the direction of cD2. This supports the notion cD2 has traveled along with the dark matter halo to the north of the gaseous subcluster. 

\subsection{Simulated Merger}
A merger scenario similar to the one described above was modeled using the FLASH hydrodynamic code \citep{Fryxell00}.  The model initial conditions are as described in \citet{Zuhone11a} for a different set of simulations with mass ratio $R=10$ and impact parameter $b\sim$1 Mpc.  When viewing the simulation in the $XY$ plane (the plane of the clusters' merger axis), the subcluster approaches on the western side of the main cluster on a southerly trajectory.  Near pericenter, it initiates sloshing in the primary cluster core.  The subcluster reaches apocenter, then falls back towards the main cluster.  Its proximity to the primary cluster's core coupled with its relative velocity causes ram pressure on its leading edge that is strong enough to remove the subcluster gas core from the subcluster potential minimum.  

\begin{figure*}[!th]
\center
\includegraphics[scale=0.83]{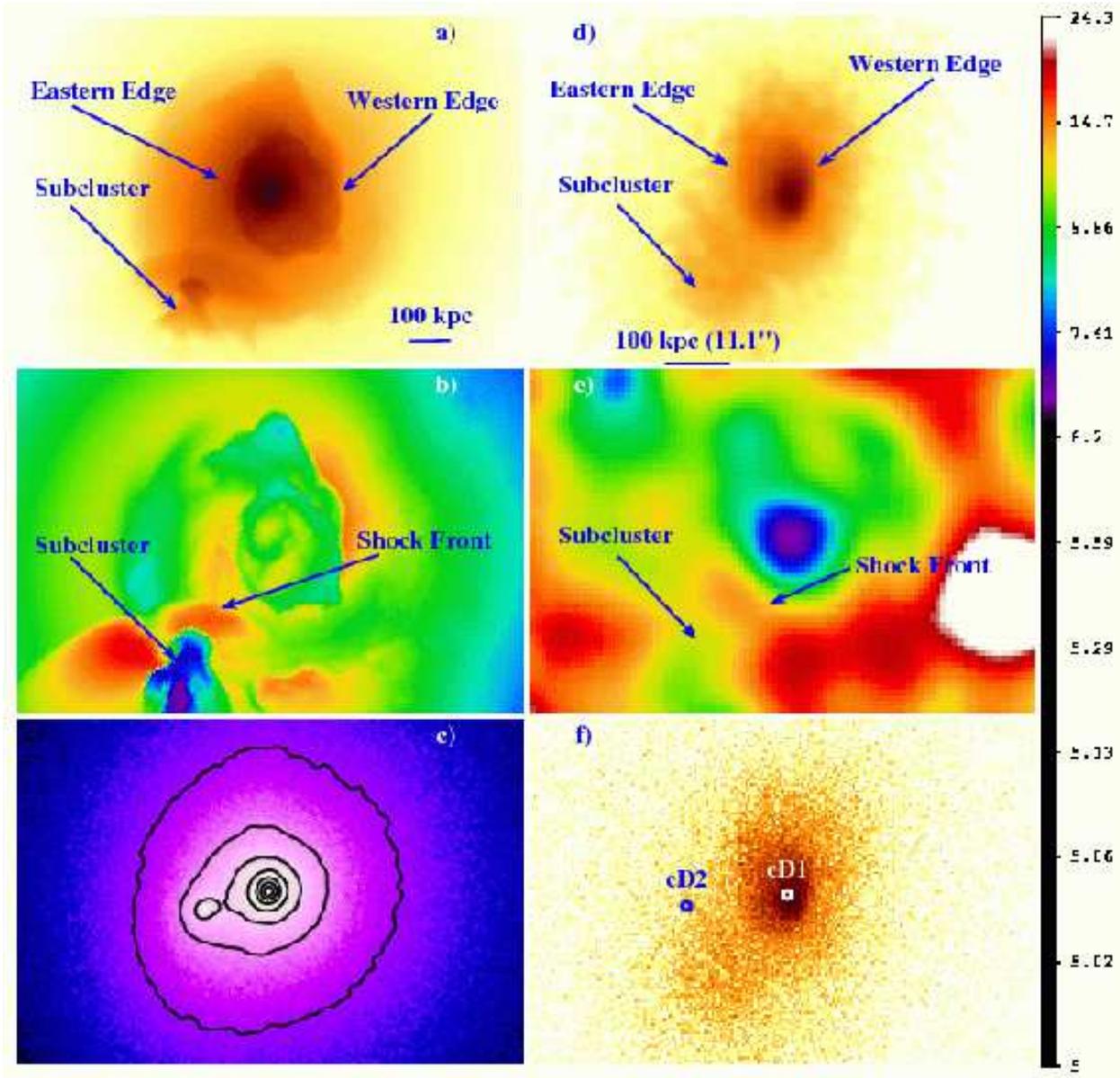}
\caption[Image comparing morphology of RXJ1347 to simulations]{\footnotesize{Figure comparing a simulated cluster merger to observations of RXJ1347.  On the left are the simulation snapshots with (a) projected surface brightness, (b) spectroscopic-like projected temperature and (c) projected mass density with linearly spaced contours of constant mass density.   The observations on the right are (d) 0.5-2.0 keV surface brightness image smooth with Gaussian (FWHM=0.5$\arcsec$), (e) temperature map as in Fig.~\ref{fig:rxj1347_f2}, (f) full resolution \textit{Chandra} surface brightness image showing the location of cD galaxies.  The spatial scales for each simulated image are the same, as are the spatial scales for each \textit{Chandra} image.  \textit{A color version of this figure is available in the online journal.}
}}
\label{fig:rxj1347_sim}
\end{figure*}

In Fig.~\ref{fig:rxj1347_sim} we show one particular snapshot in time of the simulated projected surface brightness, temperature and mass density of the cluster and subcluster and compare these with observations of RXJ1347.  Given the variety of distinct characteristics in RXJ1347 with which to compare, the qualitative similarities between the simulations and observations are remarkable.  In both the simulated and observed surface brightness images (Fig.~\ref{fig:rxj1347_sim}(a) and (d)) the sloshing edges are apparent to the east and west of the primary cluster.  The distances of the simulated fronts from the X-ray peak do appear different than those in the X-ray image.  Specifically, in the simulated image, the eastern edge is closer to the X-ray peak, indicating that the sloshing there is at a more advanced stage than in the observation.  This is not surprising given that the current location of the density peak is dependent on many factors (pericentric crossing time, impact parameter, mass ratio, the subcluster's relative velocity and orbital trajectory to name a few).  The subcluster core also appears more concentrated in the simulation image compared to the \textit{Chandra} image.  

Other merger related structures are visible when comparing the projected temperature maps (Fig.~\ref{fig:rxj1347_sim}(b) and (e)).  Despite the poorer resolution of our temperature map relative to the simulation, we can still identify the location of the sloshing edges.  We are also able to identify the cool region of the infalling subcluster.  The peak in the SZ decrement is spatially coincident with the region just between the infalling subcluster and primary cluster, however the SZ signal extends further to the north, past cD2 (see Fig.~\ref{fig:rxj1347_f2}(d)). Our merger scenario accounts for the peak in the SZ decrement, since this corresponds to the merger front between the infalling subcluster gas and the primary cluster's gas.  At this front, we observe a temperature contrast and a pressure gradient (via the SZ signal) which are both consistent with it being a shock front.  The significant SZ signal just to the north of where we identify this shock front (called the ``ridge'' in \citet{Mason10}) is not explained by our merger scenario.  It is possible that this ridge is associated with the eastern edge of the sloshing core of RXJ1347 and has no X-ray counterpart, however no such feature has been predicted in, or identified with, sloshing cores.

In our temperature map \citep[and in other works;][]{Ota08,Mason10}, there are two very hot ($\gtrsim20$ keV) regions which we do not see in the simulated temperature map.  These hot regions may have been shock heated during the first passage of the subcluster.  In the simulations, the entire cluster is heated as a result of this initial shock front, and by the second crossing the overall cluster temperature has been raised from its pre-shocked state.  The higher central temperature in the simulated cluster core is evidence of this heating.  The two hot regions, not associated with SZ decrements, which we find in our temperature map may indicate that the gas in these regions has cooled less relative to the simulation (i.e. less mixing), or that there has been some other mechanism (such as another shock front) which has maintained it at such a high temperature.

Finally, we compare the simulation projected mass density to the locations of the two cD galaxies (Fig.~\ref{fig:rxj1347_sim}(c) and (f)).  Because the galaxies and the dark matter in a merger should pass through nearly collisionlessly, the distribution of galaxies should trace the distribution of dark matter.  In particular, since a cD galaxy resides at the bottom of its cluster's potential well, it should trace the location of the dark matter density peak.  Comparing the images, we see that the location of cD2 relative to cD1 coincides with the location of the subcluster density peak relative to the main cluster density peak in the simulation.  The location of the subcluster halo in the simulation also agrees with the elongation found in lensing maps \citep{Miranda08,Bradac08}.

\section{Summary}
We have presented a new scenario for the merger history in the most luminous X-ray cluster, RXJ1347.  
This scenario is compared to optical, X-ray, and radio observations, and is able to reproduce several unique morphological features in the cluster.  We perform a new analysis on the existing X-ray data that shows the core gas is sloshing, with sloshing cold fronts present to both the east and west of the cluster core.  The morphology of the sloshing core gas places additional constraints on the merger history of the cluster: namely that the interaction which initiated the gas sloshing occurred at least 0.6 Gyr ago, and that the perturbing object first passed along the eastern side of the cluster.  In order to reconcile the gas sloshing timescale with the current positions of the nearby gaseous subcluster and second cD galaxy, the perturbing subcluster should have first passed along the western side of the primary cluster, in a southerly direction and initiated sloshing near pericenter in its orbit.  We are now seeing the subcluster when it is again near pericenter, this time heading north, where it has crossed close enough to the primary cluster's core that its gas has been ram pressure stripped.  
We make qualitative comparisons between the multiwavelength observations for the cluster and high-resolution hydrodynamic simulations.  We find that many of the unique merger structures around the cluster are well reproduced in the simulations.   This particularly well-studied cluster serves as an example of how core gas sloshing may be used to constrain the merger histories of galaxy clusters. 

\section{Acknowledgments}
The authors would like to thank the Smithsonian Institution Consortia for their support of this project.  This work has made extensive use of SAOImager DS9, in addition to software provided by the CXC in the application packages CIAO, ChIPS, and Sherpa.  This work has made use of the NASA/IPAC Extragalactic Database (NED) which is operated by the Jet Propulsion Laboratory, California Institute of Technology, under contract with the National Aeronautics and Space Administration.  The simulation software used in this work was in part developed by the DOE-supported ASC/Alliance Center for Astrophysical Thermonuclear Flashes at the University of Chicago.  R.E.J. was supported by an SAO predoctoral fellowship and a CFD fellowship during this work.  J.A.Z. was supported by {\it Chandra} grant GO8-9128X, as well as by an appointment to the NASA Postdoctoral Program at the Goddard Space Flight Center, administered by Oak Ridge Associated Universities through a contract with NASA.

\bibliographystyle{apj}                       
\bibliography{apj-jour,RXJ1347}    

\end{document}